\documentclass[showpacs,prd,aps,superscriptaddress,preprintnumbers,letterpaper]{revtex4}

\usepackage{amsmath,amssymb}
\usepackage{epsfig}
\usepackage{graphicx}
\usepackage{amsmath}
\usepackage{amsfonts}
\usepackage{epstopdf}
\usepackage{color}

\newcommand{\be}{\begin{equation}}
\newcommand{\ee}{\end{equation}}
\newcommand{\bear}{\begin{eqnarray}}
\newcommand{\eear}{\end{eqnarray}}

\newcommand{\ba}{\begin{array}}
\newcommand{\ea}{\end{array}}

\begin{document}

\title{Entropic destruction of heavy quarkonium in non-Abelian plasma \\ from 
holography}
%the holographic correspondence}%\title{Entropic dissociation: a solution for the quarkonium suppression puzzle?}

\author{Koji Hashimoto}
\affiliation{Department of Physics, Osaka University, Toyonaka, Osaka 560-0043, Japan}
\affiliation{Mathematical Physics Lab., RIKEN Nishina Center, Saitama 351-0198, Japan}
\author{Dmitri E. Kharzeev}
\affiliation{Department of Physics and Astronomy, Stony Brook University, New York 11794-3800, USA}
\affiliation{Department of Physics, Brookhaven National Laboratory, Upton, New York 11973-5000, USA}

\preprint{OU-HET-837, RIKEN-MP-95}
%%%%%%%%%
\date{\today}

\begin{abstract}
Lattice QCD indicates a large amount of entropy associated with the heavy quark-antiquark pair immersed in the quark-gluon plasma. This entropy grows as a function of the inter-quark distance giving rise to an entropic force that can be very effective in dissociating the bound quarkonium states. In addition, the lattice data show a very sharp peak in the heavy quark-antiquark entropy at the deconfinement transition. Since the quark-gluon plasma around the deconfinement transition is strongly coupled, we employ the holographic correspondence to study the entropy  associated with the heavy quark-antiquark pair in two theories: i) ${\cal{N}}=4$ supersymmetric Yang-Mills and ii) a confining Yang-Mills theory obtained by compactification on a Kaluza-Klein circle. In both cases we find the entropy growing with the inter-quark distance and evaluate the effect of the corresponding entropic forces. In the case ii), we find a sharp peak in the entropy near the deconfinement transition, in agreement with the lattice QCD results. This peak in our holographic description arises because the heavy quark pair acts as an eyewitness of the black hole formation in the bulk -- the process that describes the deconfinement transition. In terms of the boundary theory, this entropy likely emerges from the entanglement of a ``long string" connecting the quark and antiquark with the rest of the system.    
\end{abstract}


\maketitle

\setcounter{footnote}{0}

%\baselineskip 18pt \pagebreak
%\renewcommand{\thepage}{\arabic{page}}
%\tableofcontents
%\pagebreak

%\vskip0.2cm

\section{Introduction}
The studies of heavy quarkonium at finite temperature are expected to advance the understanding of QCD plasma and to clarify the nature of the deconfinement transition. It was originally proposed \cite{Matsui:1986dk} to use quarkonium suppression in heavy ion collisions as a way to detect the Debye screening in the quark-gluon plasma.  The subsequent experimental studies of quarkonium production in nuclear collisions at different energies however revealed a puzzle -- the charmonium suppression observed at RHIC \cite{Adare:2006ns} (lower energy density) appeared stronger than at LHC \cite{Abelev:2013ila} (larger energy density). This is in contrast to both the Debye screening scenario \cite{Matsui:1986dk} and the thermal activation  \cite{Kharzeev:1994pz} through the impact of gluons  \cite{Shuryak:1978ij,Bhanot:1979vb}. One possible solution to this puzzle is the recombination of the produced charm quarks into charmonia \cite{BraunMunzinger:2000px,Thews:2000rj}. 
\vskip0.2cm
However, recently it was argued \cite{Kharzeev:2014pha} that an anomalously strong suppression of charmonium near the deconfinement transition can be a consequence of the nature of deconfinement. The argument put forward in \cite{Kharzeev:2014pha} was based on the lattice QCD results \cite{Kaczmarek:2002mc,Petreczky:2004pz,Kaczmarek:2005zp,Kaczmarek:2005gi} indicating a large amount of entropy associated with the heavy quark-antiquark pair placed in the quark-gluon plasma. This entropy $S$ was found  to grow as a function of the distance $L$ between the quark and antiquark on the lattice \cite{Kaczmarek:2002mc,Petreczky:2004pz,Kaczmarek:2005zp,Kaczmarek:2005gi}. The proposal of \cite{Kharzeev:2014pha} is that this entropy should give rise to the emergent entropic force 
\be\label{entf1}
F = T\ \frac{\partial S}{\partial L} ,
\ee
where $T$ is the temperature of the plasma. It has been found that the repulsive entropic force leads to a strong suppression of charmonium states near the deconfinement transition. The leading role of the entropic force in the deconfinement transition itself has been conjectured \cite{Kharzeev:2014pha}, as well as a possible relation of the observed peak in the entropy near the deconfinement transition to the ``long string" condensation \cite{Kogut:1974ag,Pisarski:1982cn,Patel:1983sc,Aharonov:1987ah,Deo:1989bv,Levin:2004mi,Kalaydzhyan:2014tfa,Hashimoto:2014xta,Hanada:2014noa}. 
\vskip0.2cm
In this paper, we investigate the microscopic origin of the entropy associated with the heavy quark pair in non-Abelian plasma using the holographic correspondence \cite{Maldacena:1997re,Gubser:1998bc,Witten:1998qj}. We conclude that the narrow and strong peak in the entropy associated with the heavy quark pair near the transition temperature is indeed related to the nature of deconfinement, and in holographic description originates from the entropy of a long fundamental string at the bottom of the confining geometry which is absorbed into a black hole horizon at the  deconfinement transition. 
This peak is absent in the conformal ${\cal N}=4$ supersymmetric Yang-Mills theory, but emerges in a confining Yang-Mills theory obtained by compactification of the fifth dimension \cite{Witten:1998zw}. 
On the boundary, this entropy has to be attributed to long-range, delocalized excitations entangled with the heavy quark pair that can indeed be described as the ``long string". 
\vskip0.2cm
We also study the entropic force in strongly coupled gauge theories through the holographic correspondence. It turns out that the entropic force resulting from the distribution of heavy quarks in configuration space is sub-leading in the strong coupling limit, and the dynamics is dominated by the repulsive entropic force arising from the entropy of the QCD string and the attractive quark-antiquark force. The sub-leading entropic force associated with the distribution of quarks is however responsible for the real-time Einstein diffusion of quarks in the plasma once the bound state is 
dissociated; also, at realistic values of the 't Hooft coupling it does contribute to the balance of repulsive entropic and attractive potential forces and shortens the distance at which the bound states are dissociated.
\vskip0.2cm
The paper is organized as follows. In section \ref{sec:lattice} we discuss and analyze the available lattice results on the entropy of heavy quark pair in QCD plasma. In section \ref{sec:holography} we compute the entropy of the heavy quark pair in 
${\cal N}=4$ supersymmetric Yang-Mills theory and in the confining Yang-Mills theory. In section \ref{sec:entropic} we use the results for the entropy to study the entropic mechanism of quarkonium destruction. Finally, in section \ref{sec:discussion} we summarize and discuss our findings.

\begin{figure}
\includegraphics[width=15cm]{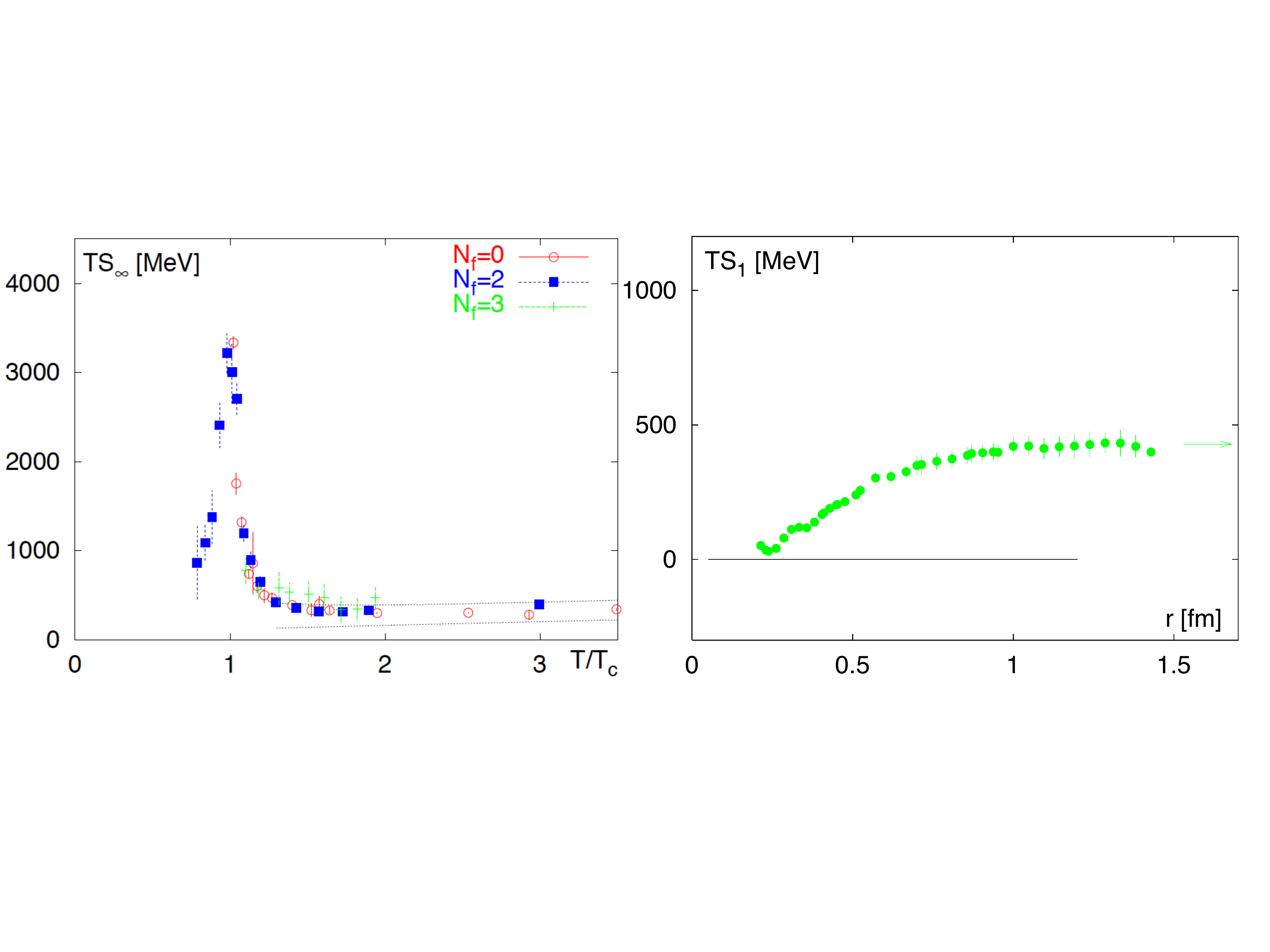}
\vspace{-3cm}
\caption{(Color online) (Left) Lattice QCD result for the entropy of the quark--antiquark pair at large separation in QCD plasma, as a function of $T/T_c$; from  \cite{Kaczmarek:2005gi}. (Right) Lattice QCD result for the entropy of the quark--antiquark pair as a function of the separation $r$ (denoted as $L$ in this paper) at temperature $T\sim 1.3\ T_c$; from  \cite{Kaczmarek:2005gi}. The arrow on the right points the value of the entropy $TS_{\infty}$ at the large distance limit, which corresponds to the left figure.}
\label{fig:lattice}
\end{figure}

\section{The entropy of the heavy quark pair in QCD plasma}\label{sec:lattice}

Lattice QCD studies indicate a large amount of additional entropy associated with the presence of static quark-antiquark pair immersed in the QCD plasma \cite{Kaczmarek:2002mc,Petreczky:2004pz,Kaczmarek:2005zp,Kaczmarek:2005gi}, see Fig. \ref{fig:lattice}. The lattice data shown in Fig. {\ref{fig:lattice}} (left) indicates the large distance limit
of the entropy associated with the quark antiquark pair. Fig.{\ref{fig:lattice}} (right) shows that this entropy grows 
as the distance $r$ between the quarks increases. The entropy growth is steep below a certain critical distance $r_c\sim 1$ fm, and then it saturates at a constant value. The entropy observed at large distances thus can be attributed
to the physics just below/around the critical distance. This dependence allows one to define the entropy $S_\infty$ at a large distance exceeding $r_c$ and study its temperature dependence -- this leads to a very pronounced peak around the transition temperature shown 
in Fig.{\ref{fig:lattice}} (left).  The value of the entropy at the peak is $S_\infty(T=T_c) \simeq 16.5$ \cite{Kaczmarek:2005gi}, corresponding to $\exp(S_\infty) \simeq 1.5 \times 10^7$ states. This large entropy is inconsistent with the picture of Debye screening based on weak coupling, as we will now discuss. 
\vskip0.2cm
Indeed, the recent lattice study of the equation of state in $(2+1)$ flavor QCD by the HotQCD Collaboration \cite{Bazavov:2014pvz} indicates that the deconfinement and chiral restoration transitions occur at the temperature $145\ {\rm MeV} < T < 163\ {\rm MeV}$, consistent with the earlier results from \cite{Aoki:2006br,Aoki:2009sc}. In the middle of the crossover region at $T_c = 155$ MeV the entropy density of the plasma is $s_c = (5.34 \pm^{0.65}_{0.62})\ T_c^3$ \cite{Bazavov:2014pvz}. The cubic box of QCD plasma of edge length $L_c \equiv 1/T_c \simeq 1.3$ fm (much larger than the screening length $L_D\simeq 0.3\ {\rm fm}$  \cite{Kaczmarek:2005ui} at this temperature) and volume $L_c^3 = T_s^{-3}$ therefore has the entropy $S_c \equiv s_c L_c^3 \simeq 5$ which is three times smaller than the additional entropy $S_\infty(T=T_c) \simeq 16.5$ \cite{Kaczmarek:2005gi} associated with the static quark-antiquark pair. 
This suggests that the presence of the heavy quark pair is felt at distances significantly exceeding the screening length $L_D$. Note that $S_\infty$ does not even include the entropy associated with different placement of heavy quarks in configuration space, and is not significantly affected by the presence of light quarks -- the entropy for different numbers  $N_f = 0, 2, 3$ of flavors is very similar, see Fig. {\ref{fig:lattice}} (Left). The entropy associated with the heavy quark pair should thus arise entirely from the additional gauge field configurations that can exist in $\exp(S_\infty) \sim 1.5 \times 10^7$ states. 
\vskip0.2cm
Let us further quantify this argument. The screening radius of the heavy quark-antiquark potential at $T_c$ is $L_D \simeq 0.3\ {\rm fm}$  \cite{Kaczmarek:2005ui}, and naively one would expect that the additional entropy due to the presence of the heavy quark pair separated by a large distance $L_\infty \gg L_D$ cannot exceed the total entropy of the plasma regions within the Debye spheres of the quark and antiquark given by $S_D \equiv 2 \times s_c\ 4 \pi L_D^3/3 \simeq 1.2$. Indeed, in the mean field treatment the quarks affect the plasma only at distances $L \leq L_D$, 
and the polarization induced by them is relatively weak and can only slightly change the entropy density of the plasma within the Debye sphere -- so the additional entropy density should be much smaller than $s_c$. Therefore, in the Debye screening picture, $S_D$ can be considered as a safe upper bound on the additional entropy induced by the presence of the heavy quark pair.   

\vskip0.2cm

The fact that the observed entropy $S_\infty(T=T_c) \simeq 16.5$ is more than an order of magnitide larger than $S_D \simeq 1.2$, $S_\infty(T=T_c) \gg S_D$ indicates that in the transition region around $T=T_c$ the heavy quark pair is entangled with the QCD plasma at distances that far exceed the screening radius of the potential. A possible way to accommodate this observation is to assume that the heavy quark pair around $T=T_c$ is connected by a ``long string"  that does not contribute to the internal quark-antiquark energies at distances exceeding the screening length $L>L_D$, but does contribute to the entropy of the system. 

\vskip0.2cm

At the very least, the lattice data indicate a dramatic breakdown of the weak coupling, mean field approach to heavy quark screening near the deconfinement transition region and calls for a use of a method valid at strong coupling. Because of this, we will rely on the holographic gauge/gravity correspondence and investigate the entropy associated with the heavy quark pair within two different theories: ${\cal N}=4$ supersymmetric Yang-Mills (SYM) theory that at $T=0$ possesses conformal invariance, 
and pure Yang-Mills (YM) theory that possesses confinement at low temperatures and the deconfinement phase transition  \cite{Witten:1998zw}. 

\section{The entropy of the heavy quark pair from holography}
\label{sec:holography}

Using the AdS/CFT correspondence \cite{Maldacena:1997re,Gubser:1998bc,Witten:1998qj},
we will now evaluate the additional entropy associated with adding a color singlet heavy quark-antiquark pair to the non-Abelian plasma.
In this section, we demonstrate the emergence of the entropy using  
two examples: {\it i)} ${\cal N}=4$ supersymmetric Yang-Mills (SYM) theory
and {\it ii)} pure Yang-Mills (YM) theory \cite{Witten:1998zw}. 
The latter is realized only as a low energy limit,
and our high temperature phase corresponds to a SYM theory in 5 spacetime dimensions
compactified on a circle of radius $1/M_{\rm KK}$. 

\subsection{The entropy of quark pair in ${\cal N}=4$ supersymmetric Yang-Mills theory}

\begin{figure}
\includegraphics[width=6.5cm]{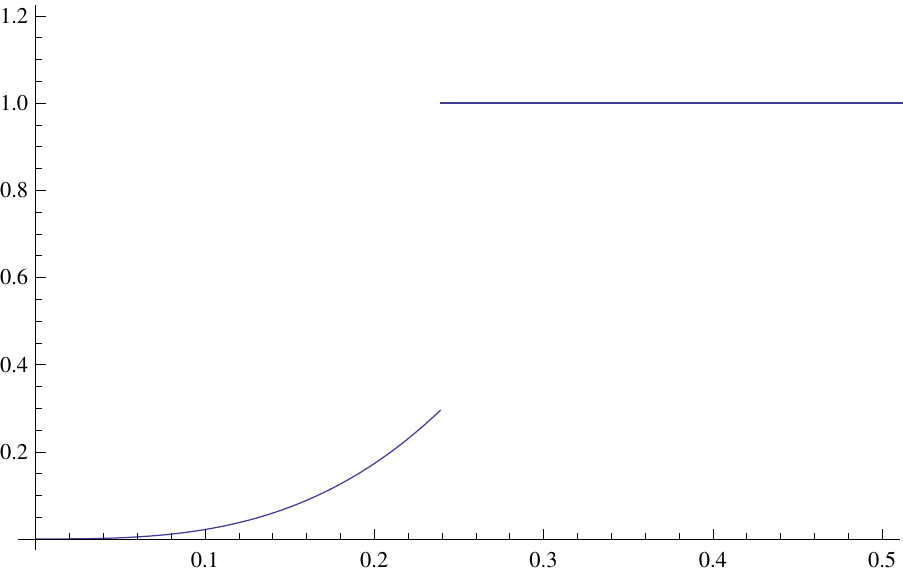}
\put(10,5){$LT$}
\put(-200,120){$S_{\rm str}/\sqrt{2\lambda}$}
\put(-160,30){$S_{\rm str}^{(1)}/\sqrt{2\lambda}$}
\put(-70,110){$S_{\rm str}^{(2)}/\sqrt{2\lambda}$}
\caption{(Color online) The entropy of the quark-antiquark pair in the ${\cal N}=4$ SYM plasma, as a function of $LT$. For $LT>c\simeq 0.24$, the quark and the antiquark are screened and the
entropy is constant ($S_{\rm str}^{(2)}$). For smaller values of $LT$, the entropy grows approximately as $(LT)^3$ (see the expression (\ref{S2}) for  $S_{\rm str}^{(1)}$).}
\label{fig:entropy}
\end{figure}
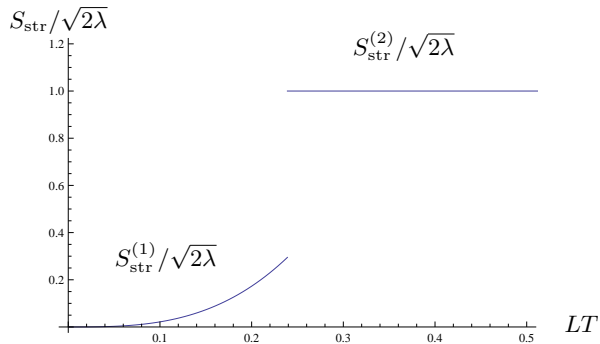

The calculation of the quark-antiquark potential at zero temperature 
in the ${\cal N}=4$ supersymmetric Yang-Mills (SYM) theory was 
given in \cite{Maldacena:1998im,Rey:1998ik}, and was extended to
the finite temperature case in \cite{Rey:1998bq}. We follow the latter approach 
to derive the entropic force acting on a quark-antiquark pair in a hot ${\cal N}=4$ SYM plasma.
\vskip0.2cm
In the AdS/CFT correspondence, the free energy of a quark-antiquark pair is equal to
an on-shell action of the fundamental string in the dual gravity geometry, with asymptotic separation $L$ 
taken to be equal to the inter-quark distance. The $AdS_5$ Schwarzschild black hole 
geometry is
\be
ds^2 = \frac{r^2}{R^2} \left(-f(r) dt^2 + d\vec{x}^2 \right)
+ \frac{R^2}{r^2} f(r)^{-1} dr^2 + R^2 d\Omega_5^2
\ee
where
$f(r) \equiv 1-r_T^4/r^4$ and $r_T\equiv \pi R^2 T$, with a temperature $T$ of the gauge theory.
$R$ is the AdS radius that is related to the gauge theory coupling constant as $R^4 =2 l_s^4 \lambda$
where $\lambda \equiv g_{\rm YM}^2 N_c$ is the 't Hooft coupling, and $N_c$ is the number of colors.
We follow the calculation of \cite{Rey:1998bq} and obtain the on-shell action of the 
fundamental string as
\be
F_{\rm str}^{(1)} = \frac{\sqrt{2\lambda}}{\pi} \int_{U_{\rm min}}^\infty dU 
\sqrt{\frac{U^4-(\pi T)^4}{U^4-U_{\rm min}^4}}. 
\label{F1}
\ee
We have redefined the radial coordinate in the AdS as $U\equiv r/R^2$. 
The string is bent down toward the black hole horizon, and is U-shaped. $U_{\rm min}$ is 
the lowest position of the fundamental string in the AdS Schwarzschild geometry. For a given 
$U_{\rm min}$, the inter-quark distance can be calculated by integrating the equation of motion of
the fundamental string, as
\be
L = 2\int_{U_{\rm min}}^\infty dU 
\sqrt{\frac{U_{\rm min}^4-(\pi T)^4}{(U^4-(\pi T)^4)(U^4-U_{\rm min}^4)}}.
\label{LUmin}
\ee
Eliminating $U_{\rm min}$ in (\ref{F1}) and (\ref{LUmin}), we obtain the quark-antiquark 
free energy 
as a function of $L$ and $T$.
\vskip0.2cm
If $L$ is large enough, the fundamental string breaks in two pieces and the quarks are screened. The free energy for this case is given as
\be
F_{\rm str}^{(2)} =\frac{\sqrt{2\lambda}}{\pi} \int_{\pi T}^\infty dU,
\label{F2}
\ee
which is nothing but the proper length of two fundamental strings extending 
from the black hole horizon to the AdS boundary. Comparing the two free energies $F_{\rm str}^{(1)}$ and $F_{\rm str}^{(2)}$,
we find that quarks are completely screened ($F_{\rm str}^{(2)}$ is favored) 
if $L> c/T$ where numerically $c \simeq 0.240$.
\vskip0.2cm
Let us calculate the entropy $S = -\partial F/\partial T$. For the screened $L>c/T$, we easily obtain 
\be
S_{\rm str}^{(2)} = \sqrt{2\lambda} \, \theta(L-c/T).
\label{S1}
\ee
To evaluate the entropy for $L<c/T$, we rewrite (\ref{LUmin}) with $u \equiv U/T$ ($u_{\rm min}\equiv U_{\rm min}/T$) and  $s \equiv u/u_{\rm min}$ as
\be
L = \frac{2}{u_{\rm min}T} \int_1^\infty ds \sqrt{\frac{1-(\pi/u_{\rm min})^4}{(s^4-(\pi/u_{\rm min})^4)(s^4-1)}}.
\ee
Numerically, for the region of our interest $L<c/T$, this expression can be approximated by ignoring
the ``horizon effect" $(\pi/u_{\rm min})^4$ within a few percent error as
\be
L \simeq 
\frac{2}{u_{\rm min}T} \int_1^\infty ds \sqrt{\frac{1}{s^4(s^4-1)}} = \frac{d}{u_{\rm min}T},
\ee
where $d \equiv 2 \sqrt{\pi}\Gamma(3/4)/\Gamma(1/4)\simeq 1.20$. 
Using this, the free energy is
\bear
F_{\rm str}^{(1)} &=& \frac{\sqrt{2\lambda}}{\pi} T \int_1^\infty ds \; 
u_{\rm min} \sqrt{\frac{s^4-(\pi/u_{\rm min})^4}{s^4-1}}
\nonumber \\
&\simeq &
\frac{\sqrt{2\lambda}d }{\pi L}  \int_1^\infty ds \sqrt{\frac{s^4-(\pi/u_{\rm min})^4}{s^4-1}}.
\eear
Then, the entropy is
\be
S_{\rm str}^{(1)} = -\frac{\partial F_{\rm str}^{(1)}}{\partial T}
 \simeq
 2\sqrt{s\lambda} \left(\frac{\pi LT}{d}\right)^3
 \int_1^\infty ds \sqrt{\frac{1}{(s^4-(\pi/u_{\rm min})^4)(s^4-1)}}.
\ee
Again, ignoring the ``horizon effect", this can be approximated by
\be
S_{\rm str}^{(1)}
 \simeq
 2\sqrt{s\lambda} \left(\frac{\pi LT}{d}\right)^3
 \int_1^\infty ds \sqrt{\frac{1}{s^4(s^4-1)}} = \sqrt{2\lambda}\ \frac{\pi^3}{d^2} \ (LT)^3.
  \qquad (L<0.24/T)
 \label{S2}
\ee
This is the approximate expression for the entropy of the quark antiquark pair for $L<c/T$.
Together with (\ref{S1}), the behavior of the entropy is shown in Fig.\ref{fig:entropy}. While we qualitatively reproduce the growth of the entropy with the inter-quark distance, the theory does not possess a deconfinement transition, and the entropy at large distance does not depend on the temperature. To be able to address the behavior around the deconfinement transition, we need a theory which at $T=0$ possesses confinement.   

%%%%%%%%%%%%%%%%%%%%%%%
%%%%%%%%%%%%%%%%%%%%%%%

\subsection{The entropy of heavy quark pair in a confining Yang-Mills theory}

A well-known way to obtain a pure Yang-Mills theory in 4 dimensions in AdS/CFT is
to start with a SYM in 5 dimensions and compactify it on a Kaluza-Klein (KK) circle
$x^4 \sim x^4 + 2 \pi/M_{\rm KK}$ with anti-periodic boundary condition for fermions 
\cite{Witten:1998zw}. In the low energy limit the theory is expected to be identical to the pure Yang-Mills theory. At high temperatures,  the theory is in a deconfined phase
with a dual gravity metric
\be
ds^2 = \left(\frac{U}{R}\right)^{3/2}
(-f(U) dt^2 + d\vec{x}^2 + dx_4^2)
+\left(\frac{R}{U}\right)^{3/2} \left(\frac{1}{f(U)} dU^2 + U^2 d\Omega_4^2\right).
\ee
This metric is nothing but a black D4-brane with temperature \footnote{Another proposal for 
a high temperature phase can be found in \cite{Mandal:2011ws}.}
\be
T = \frac{3}{4\pi} \frac{U_{\rm KK}^{1/2}}{R^{3/2}},
\ee
where $U=U_{\rm KK}$ is the black brane horizon with $f(U)\equiv 1-(U_{\rm KK}/U)^3$,
and $R$ is related to the gauge coupling constant in 4 dimensions as
$\lambda = 2R^3 M_{\rm KK}/l_s^2$.
The free energy of a fundamental string in this geometry was calculated in \cite{Rey:1998bq}, 
and we follow that computation. At $L<c'/T$ with $c'\sim 0.278$, the quark is not screened when seen from the antiquark, and we obtain
\bear
F_{\rm str}^{(1)} &=& \frac{\lambda T^2}{2\pi M_{\rm KK}}
\int_1^\infty ds \ u_{\rm min} \sqrt{\frac{s^3-(16 \pi^2 /3u_{\rm min})^3}{s^3-1}} ,
\\
L &=& \frac{2}{T} \int_1^\infty ds \ u_{\rm min}^{-1/2}
\sqrt{\frac{1-(16 \pi^2 /3u_{\rm min})^3}{(s^3-(16 \pi^2 /3u_{\rm min})^3)(s^3-1)}} . 
\eear
Here we have made a redefinition $U = s u_{\rm min} R^3 T^2$.
For the screened $L > c'/T$, the free energy is
\be
F_{\rm str}^{(2)} = \frac{\lambda}{2\pi M_{\rm KK} R^3} \int_{(4 \pi T/3R^3)}^\infty dU.
\label{F2prime}
\ee
The entropy can be calculated in the same manner, within the same approximation. We find
\be\label{Ldep}
S_{\rm str}^{(1)} \simeq \frac{2^7 \pi^5}{3^5 d'^3}\frac{\lambda}{M_{\rm KK}} L^4 T^5
 \qquad (L<0.28/T)
\ee
where $d' \equiv \int_1^\infty ds \ (s^3 (s^3-1))^{-1/2} \simeq 0.86$.
On the other hand, for $L>c'/T$, we again find a constant entropy,
\be
S_{\rm str}^{(2)} = \frac{16\pi}{9} \frac{\lambda}{M_{\rm KK}} T. \qquad (L>0.28/T)
\label{S2con}
\ee
The behavior of the entropy as a function of $L$ looks quite similar to the case of the
${\cal N}=4$ SYM in Fig.~\ref{fig:entropy} -- the only difference is a larger power of $L$
for $L<c'/T$.
\vskip0.2cm

\begin{figure}
\includegraphics[width=7cm]{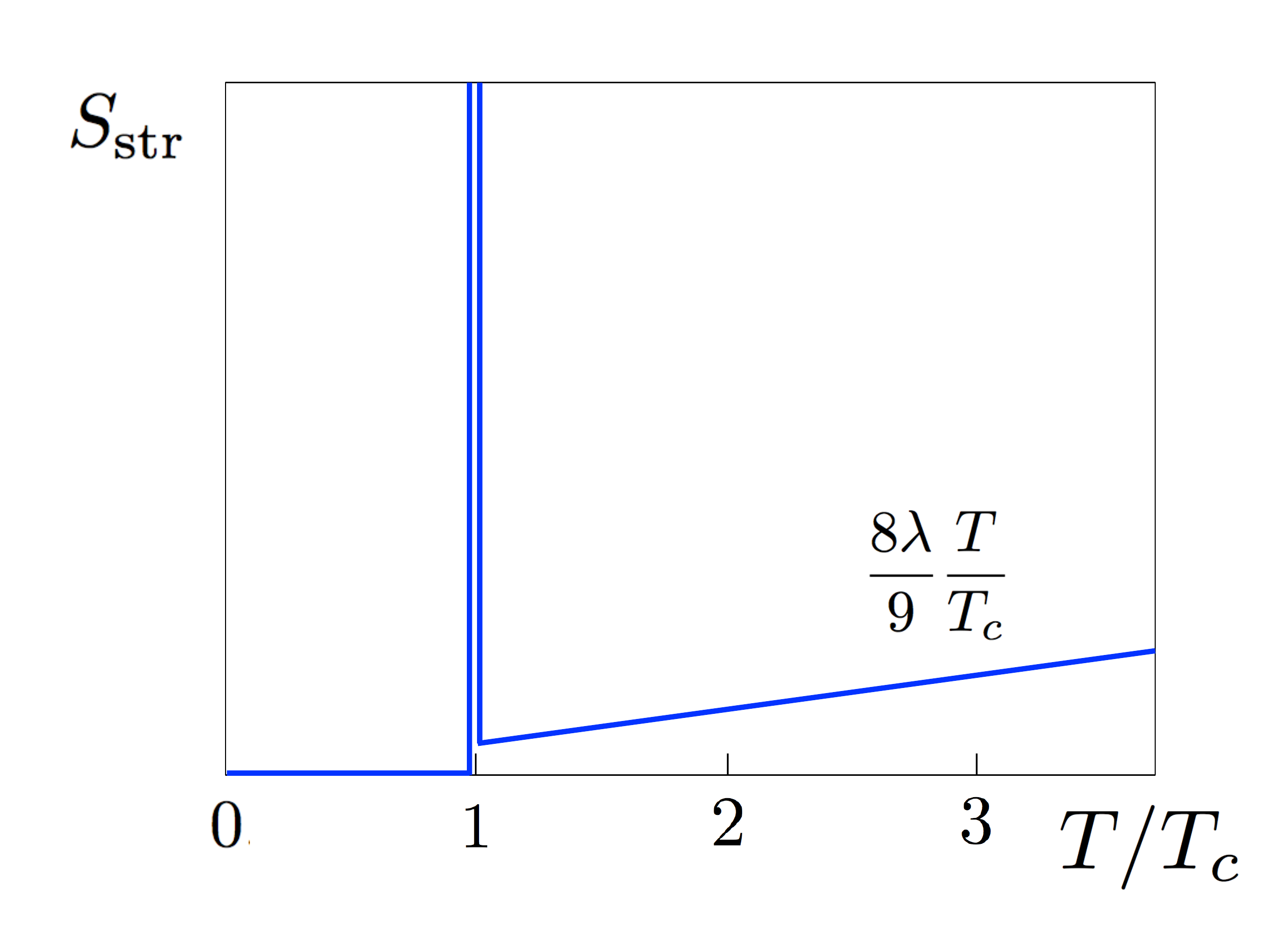}
\caption{
(Color online) A schematic plot of the entropy of the quark-antiquark pair at large distance in the confining Yang-Mills theory as a function of $T/T_c$, as given by the holographic result (\ref{entroform}). 
The entropy diverges at $T=T_c$. For low temperatures $T<T_c$ the entropy vanishes, while for high temperatures $T>T_c$ it has a finite value proportional to the temperature.
}
\label{fig:entropy2}
\end{figure}

Let us now check how the entropy at large $L$ depends on the temperature. 
The deconfinement transition in the gauge/gravity model occurs at 
the critical temperature $T_{\rm c}=M_{\rm KK}/2\pi$. For $T<T_{\rm c}$,  
the geometry does not depend on the temperature. So the free energy of the quark antiquark pair, which is the free energy of 
the fundamental string, is independent of the temperature, and is 
proportional to $L$ at large $L=L_{\infty}$ because of the confining geometry,
\be
F_{\rm str} \sim  \sigma_{\rm QCD} L_\infty
+\frac{\lambda}{2\pi M_{\rm KK} R^3} \int_{(2/(3R^3M_{\rm KK}))}^\infty dU,
\label{largeL}
\ee
where $\sigma_{\rm QCD}$ in the first term is the QCD string tension as computed in the gauge/gravity model,
\be
\sigma_{\rm QCD}=\frac{2\lambda}{27\pi}M_{\rm KK}^2 .
\ee
The second term is $F^{(2)}_{\rm str}$ at $T=T_{\rm c}$. The approximation employed here is that the string shape is almost rectangular, which is a good approximation for 
large $L=L_{\rm \infty}$.
In the confining phase at $T<T_{\rm c}$ the free energy does not depend on $T$, so the entropy vanishes,
\be
S_{\rm str} =0 \quad (T<T_{\rm c})
\ee
On the other hand, in the deconfined phase at $T>T_{\rm c}$, the entropy is nonzero and is given by Eq.~(\ref{S2con}).
\vskip0.2cm
When the temperature approaches the critical temperature $T_{\rm c}$, the free energy jumps from the value (\ref{largeL}) to the value (\ref{F2}). Therefore we get an entropy given by the delta function at $T=T_{\rm c}$
with a positive coefficient that reflects the jump in the free energy.
To summarize, the entropy at large $L=L_\infty$ is given by 
\be
S_{\rm str} = \sigma_{\rm QCD} L_\infty \, \delta (T-M_{\rm KK}/2\pi)+ \frac{16 \pi}{9}\frac{\lambda}{M_{\rm KK}} T \, \, \theta(T-M_{\rm KK}/2\pi)\, .
\label{entroform}
\ee
A schematic plot of this entropy as a function of the temperature is shown in Fig.\ref{fig:entropy2}.

\vskip0.2cm

In the real QCD, the deconfinement transition is a cross-over,  and so the
delta-function will be smeared. Furthermore, the model we consider is not a good approximation to QCD at high temperature
as the gauge/gravity model is not asymptotically free, so the behavior at very high $T$ cannot be trusted. Keeping these simplifications of the model in mind, we may say that the entropy formula (\ref{entroform}) captures the characteristic behavior of the entropy found in lattice QCD \cite{Kaczmarek:2005gi}, Fig.\ref{fig:lattice} (left), and most importantly reproduces the sharp peak at the transition temperature. We will make a more detailed comparison to the lattice data in the next section.

%%%%%%%%%%%%%%%%%%%%%%%
%%%%%%%%%%%%%%%%%%%%%%%

\vskip0.2cm

\begin{figure}
\includegraphics[width=15cm]{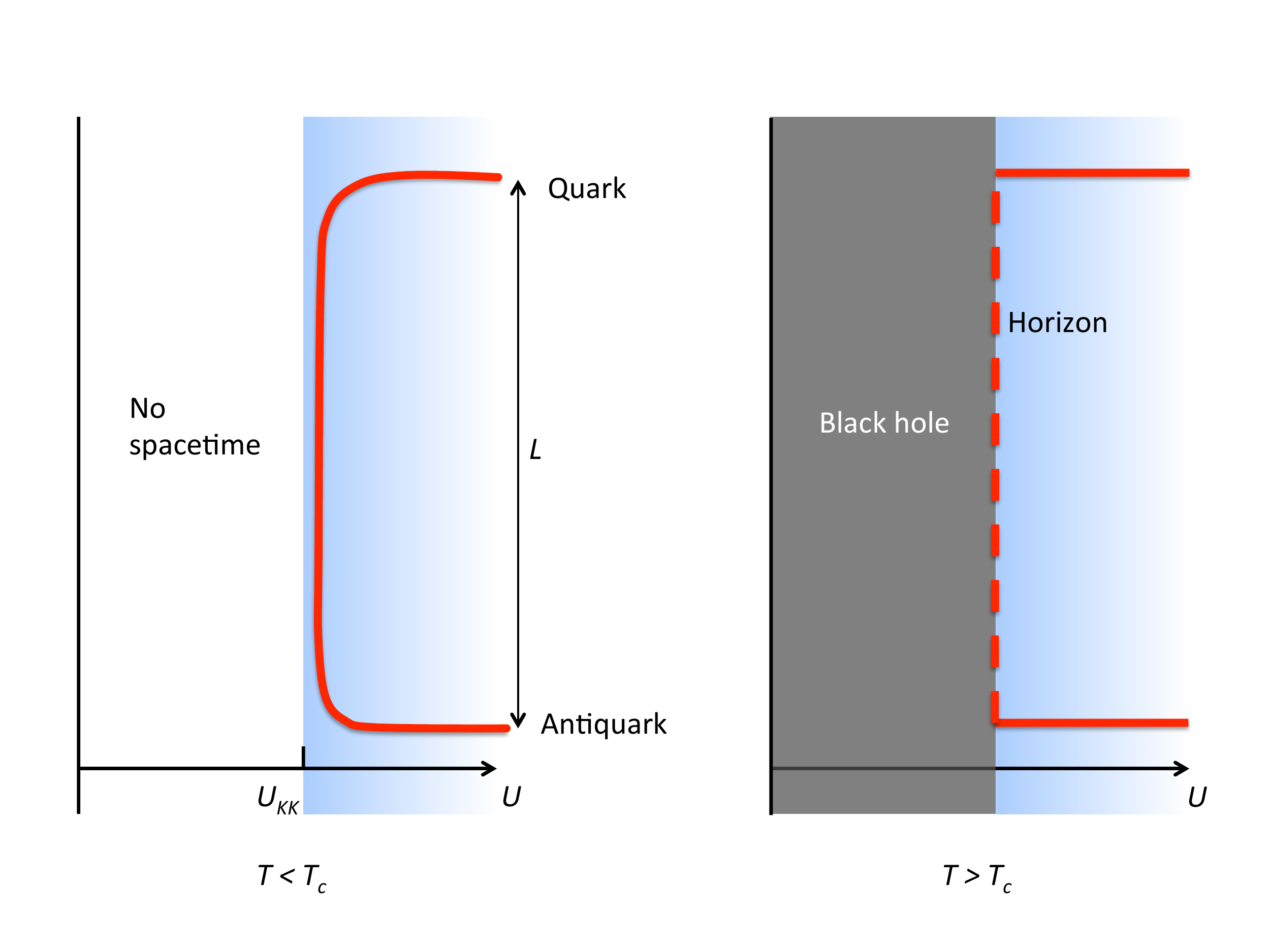}
\caption{
(Color online) A sketch of the holographic dual of the quark-antiquark pair in the confining gauge theory. The left panel depicts a confining phase, and the right panel -- the deconfined phase.  
The horizontal axis is the holographic direction $U$, and the vertical axis is our 3-dimensional space. 
The quark-antiquark pair is connected by a fundamental string hung down from the boundary of the geometry (located at the right edge of each figure).
(Left) At low temperatures, and at a large inter-quark distance $L$, the string creeps at the bottom of the geometry. (Right) For a temperature higher than the critical temperature $T_c$, the string is divided into two parts. The dashed line is the would-be string at the horizon just at $T=T_c$ resulting in the peak of the entropy associated with the quark-antiquark pair.}

\label{fig:horizon}
\end{figure}

The sudden peak of the entropy at $T=T_c$ from the holographic viewpoint can be attributed to the process of black hole formation as detected by the quark-antiquark pair. Indeed, 
suppose we gradually increase the temperature from $T=0$ in the presence of the very long fundamental string connecting the quark and antiquark at large distances. The string creeps at the bottom of the confining geometry. The
long string can fluctuate but this costs a large energy comparable to the confining scale (QCD string tension at zero temperature). Now, when the temperature reaches the critical temperature $T_c$, the background geometry is suddenly replaced by 
that with a large black hole. Typically the location of the black hole horizon is exactly at the same place where the bottom of the confining geometry used to be because it is a Hawking-Page transition --- so the long fundamental string is placed at the horizon, see Fig.\ref{fig:horizon}. The string at the horizon has a vanishing effective tension because of the red shift, and the fluctuation does not cost any energy -- thus the effective temperature becomes infinite and the entropy diverges. This simple argument may thus explain the peak of the quark pair entropy at $T=T_c$. The peak observed in the lattice data in Fig.\ref{fig:lattice} (left) from the holographic viewpoint thus can be interpreted as reflecting the formation of the black hole horizon.

%%%%%%%%%%%%%%%%%%%%%%%
%%%%%%%%%%%%%%%%%%%%%%%

%%%%%%%%%%%%%%%%%%%%%%%%%%%%%%%%%%%%%%%%%
\section{Comparison to lattice QCD, and entropic destruction of quarkonium}\label{sec:entropic}

In the previous sections we established the existence of the entropy associated with the heavy quark pair both in 
i) ${\cal{N}}=4$ supersymmetric Yang-Mills and in ii) the confining Yang-Mills theories. In both cases the entropy first increases as a function of the inter-quark distance $L$, and then becomes independent of $L$. These qualitative features are in agreement with the lattice observations \cite{Kaczmarek:2005gi}, and lead to the entropic force inducing quarkonium dissociation \cite{Kharzeev:2014pha}. However the dependence of the entropy $S_\infty$ at large inter-quark distance on the temperature is very different in the considered theories
i) and ii). In ${\cal{N}}=4$ supersymmetric Yang-Mills we get a temperature-independent value $S_\infty = \sqrt{2 \lambda}$, see (\ref{S1}). 
This is in sharp contrast to the lattice results, see Fig. \ref{fig:lattice} (Left), that show a sharp peak in $S_\infty$ near the deconfinement transition temperature. 
\vskip0.2cm
 On the other hand, the holographic confining Yang-Mills theory \cite{Witten:1998zw} produces the entropy (\ref{entroform}) that has a sharp delta-function peak around the deconfinement transition, in a qualitative agreement with the lattice data. Let us try to make a semi-quantitative comparison of our holographic result (\ref{entroform}) with the data. We choose to measure the entropy at the distance $L_\infty= 1/T_c$ which is already in the screening regime (\ref{F2}) of the free energy; this is also consistent with the lattice results, see Fig. \ref{fig:lattice} (right) that shows the flattening of the entropy at $L \simeq 1$ fm = $1/T_c$ for $T_c = 200$ MeV as for $N_f =2$ QCD \cite{Kaczmarek:2005gi}.
The integral over the delta-function peak then yields 
 \be
\int S_{\infty}\ dT = \sigma_{\rm QCD} L_\infty = \frac{\sigma_{\rm QCD}}{T_c}. 
 \ee
 Tuning the coupling $\lambda$ of our theory to describe the physical QCD string tension of $\sigma_{\rm QCD} \simeq 0.8$ GeV/fm and using $T_c = 200$ MeV as for $N_f =2$ QCD \cite{Kaczmarek:2005gi} we get the value
 $\int S_{\infty}\ dT \simeq 0.8$ GeV that is to be compared to the lattice data. 

\vskip0.2cm

 We can estimate the integral over the entropy peak in the lattice data as  $\int S^{\rm lat}_{\infty}\ dT \simeq S_{\rm peak} (T_u - T_l)$, where $S_{\rm peak} = 16.5$ \cite{Kaczmarek:2005gi} is the entropy at $T=T_c$, and $T_u$ and $T_l$ are the temperatures above and below $T_c$ at which the entropy is equal to $0.5\  S_{\rm peak}$. The paper \cite{Kaczmarek:2005gi} gives the values of $T_u = 1.07\ T_c$ and $T_l = 0.89\ T_c$ leading to $\int S^{\rm lat}_{\infty}\ dT \simeq 0.18\ T_c\ S_{\rm peak} \simeq 0.6$ GeV, where as before we used $T_c = 200$ MeV for $N_f =2$ QCD \cite{Kaczmarek:2005gi}. This value of the integral is reasonably close to the prediction of the holographic model, $\int S_{\infty}\ dT \simeq 0.8$ GeV.
 We thus conclude that our holographic approach provides an adequate description of the lattice results in the transition region. 

%%%%%%%%%
\vskip0.2cm
Let us now discuss the dependence of the entropy on the inter-quark distance $L$. The lattice data (see Fig. \ref{fig:lattice} (right)) shows that the entropy grows with $L$, and then saturates at some critical distance. We can compare this behavior of the lattice data to our holographic predictions from supersymmetric Yang-Mills theory shown in Fig.\ref{fig:entropy}. One can observe that the gross features of the $L$-dependence -- increase with $L$ followed by a saturation at some critical distance -- are captured by our approach, even though the detailed shape is not reproduced. In the confining Yang-Mills theory at 
temperatures above the deconfinement transition the dependence of the entropy on inter-quark distance is similar, with a somewhat steeper growth at small $L$, see (\ref{Ldep}). The growth of the entropy with the distance in our approach is responsible for the entropic force (\ref{entf1}) that tends to destroy the bound quark-antiquark states. 
\vskip0.2cm
We can now examine the value of the entropy at large distance.  The lattice value at $T=1.3\ T_c$ (shown by the arrow on Fig. \ref{fig:lattice}, right) is $S_\infty =1.63$ \cite{Kaczmarek:2005gi}. The large distance limit of the entropy in supersymmetric Yang-Mills theory at large $T$ is given holographically by (\ref{S1}), which is
$S_\infty=\sqrt{2\lambda}$. Let us choose the value of the coupling $\lambda = 5.5$ determined \cite{Gubser:2006qh} by fitting the quark-antiquark short-distance Coulomb force to the lattice data. Then the holographic estimate for ${\cal N}=4$ SYM is $S_\infty \simeq 3.3$, within a factor of two from the lattice QCD result. In the confining Yang-Mills theory, the value of the entropy at large distance is given by (\ref{entroform}) with $T_c = M_{KK}/2\pi$: $S_\infty = (8 \lambda/9)\ T/T_c$. Using the same value of the coupling $\lambda = 5.5$ at $T=1.3\ T_c$ we get an estimate $S_\infty \simeq 6.3$, within a factor of four from the lattice QCD result. Note that the agreement between the lattice QCD and the confining holographic Yang-Mills worsens at high temperatures, possibly reflecting the lack of asymptotic freedom in the latter theory.

 \vskip0.2cm

%%%%%%%
Now, basing on the holographic calculation of the quark-antiquark entropy and the corresponding entropic force (\ref{entf1}), we can evaluate the critical distance at which the heavy quark-antiquark pair is to be destructed.
Let us consider first the short distance regime $L<0.24/T$ in which the entropy is given by (\ref{S2}): 
\be\label{entropy_string}
S_{\rm str} = \sqrt{2 \lambda}\ \frac{\pi^3}{d^2}\ (L T)^3,
\ee
where $d \equiv 2 \sqrt{\pi}\Gamma(3/4)/\Gamma(1/4)\simeq 1.20$. 
The entropic force (\ref{entf1}) is then 
\be\label{string_ent}
F_{\rm str} = T \frac{\partial S}{\partial L} =  \sqrt{2 \lambda}\ \frac{3 \pi^3 T^4}{d^2}\ L^2 .
\ee
The concept of the entropic force relies on the assumption that the number of interactions needed to change $L$ substantially is very large. The ensemble average is thus performed over the continuous three-dimensional Gaussian probability distribution 
\be\label{prob_dist}
P(L) = \frac{4\ L^2}{\sqrt{\pi}\ q(t)^3} \ \exp\left(-\frac{L^2}{q(t)^2}\right) ,
\ee 
defined as follows: after time $t$ the particle will be located between $L$ and $L + dL$ with the probability $P(L) dL$ normalized by $\int P(L) dL = 1$, and $q(t)$ is the most probable value of $r(L)$. It is well known that the Gaussian distribution as a limit of Bernoullian distributions when the number of steps in a walk becomes very large \cite{Chandra}. 
\vskip0.2cm
Performing the Gaussian ensemble averaging of (\ref{string_ent}), we get
\be\label{string_force}
\langle F_{\rm str} \rangle = \sqrt{2 \lambda}\ \frac{9 \pi^3 T^4}{2 d^2}\ q^2 .
\ee
In addition, we have to consider the entropic force due to the $L$-dependence of the number of states. 
Consider a particle released at the origin $L=0$. The number of states for the particle at distances between $L$ 
and $L + dL$ is proportional to the volume $dV(L) = 4 \pi L^2 dL \equiv \Omega(L) dL$, and the corresponding $L$-dependent part of the entropy is 
\be\label{entdif}
S_{\rm diff}(L) = \ln \Omega(L) = 2 k  \ln L + {\rm const} ;
\ee
where we put the Boltzmann constant $k = 1$.
The resulting entropic force is 
\be\label{entf}
F_{\rm diff}(L) = T \frac{\partial S}{\partial L} = \frac{2 k T}{L} .
\ee
Perfoming the Gaussian average of (\ref{entf}), we get
\be\label{dif_force}
\langle F_{\rm diff} \rangle = \frac{4 T}{\sqrt{\pi} q} .
\ee
Both entropic forces considered above point outward, since both (\ref{entropy_string}) and (\ref{entdif}) grow when the distance $L$ between the quark and antiquark increases. 
\vskip0.2cm
These entropic forces are balanced by the attractive quark-antiquark potential which in ${\cal{N}}=4$ supersymmetric Yang-Mills theory is of Coulomb type and is given by \cite{Maldacena:1998im,Rey:1998ik} 
\be\label{coul}
U_C(L) = - \frac{4 \pi^2 \sqrt{2 \lambda}}{(\Gamma\left(\frac{1}{4}\right))^4}\ \frac{1}{L} ,
\ee
with the corresponding force
\be\label{coul_force}
F_C = - \frac{\partial U_C}{\partial L} = - \frac{4 \pi^2 \sqrt{2 \lambda}}{(\Gamma\left(\frac{1}{4}\right))^4}\ \frac{1}{L^2} .
\ee
At short distances $L$ that we consider the temperature dependence of this potential is weak and can be neglected for our estimate.
The Gaussian ensemble average of (\ref{coul_force}) yields
\be\label{coul_ave}
\langle F_C \rangle = - \frac{8 \pi^2 \sqrt{2 \lambda}}{(\Gamma\left(\frac{1}{4}\right))^4}\ \frac{1}{q^2} .
\ee
Balancing the attractive Coulomb force (\ref{coul_ave}) against the entropic forces (\ref{string_force}) and (\ref{dif_force}), 
\be
- \langle F_C \rangle = \langle F_{\rm str} \rangle + \langle F_{\rm diff} \rangle,
\ee
we get
\be\label{balance}
\frac{8 \pi^2 \sqrt{2 \lambda}}{(\Gamma\left(\frac{1}{4}\right))^4}\ \frac{1}{q^2} = \sqrt{2 \lambda}\ \frac{9 \pi^3 T^4}{2 d^2}\ q^2 + \frac{4 T}{\sqrt{\pi} q} .
\ee
Let us first consider the solution in the strong coupling limit $\lambda \to \infty$ appropriate for the classical gravity approximation in the bulk. In this limit we can neglect the last term in (\ref{balance}); introducing a dimensionless variable $t \equiv q T$, we find the equilibrium value of $t=t_0$:
\be\label{t_strong}
t_0^2 = \frac{4}{3\sqrt{\pi}}  \frac{d}{(\Gamma\left(\frac{1}{4}\right))^2} ;
\ee
note that this expression is independent of the coupling $\lambda$. 
Since the inter-quark distance $x$ in three spatial dimensions is related to $q$ by $q^2 = 2 x^2$, we find from (\ref{t_strong}) 
\be\label{t_strong1}
\langle x_0^2 \rangle = \frac{2}{3\sqrt{\pi}}  \frac{d}{(\Gamma\left(\frac{1}{4}\right))^2} \ T^{-2} ,
\ee
which is our final result for the strong coupling ${\cal{N}}=4$ supersymmetric Yang-Mills case. Numerically, we find $\langle x_0^2 \rangle \simeq 0.034\ T^{-2}$; at temperature $T = 200$ MeV (which is the deconfinement transition temperature in two-flavor QCD) we thus get  
\be\label{est_strong}
\bar{x}_0 (T = 200\ {\rm MeV}) \equiv  \sqrt{\langle x_0^2 \rangle} \simeq 0.18\ {\rm fm},
\ee
which is somewhat below the size of the ground state of charmonium. It is interesting to note that this distance scale emerges from the prefactor of $T^{-2}$ in (\ref{t_strong1}) which is a pure number.
\vskip0.2cm
The fact that $\bar{x}_0(T) \sim T^{-1}$ is of course a consequence of conformal invariance of the theory -- $T$ is the only dimensionful scale in the problem. In this conformal case, (\ref{balance}) suggests the following interpretation of the scale $\bar{x}_0(T)$. At $x < \bar{x}_0(T)$, the Coulomb force dominates, and the appropriate formulation of the 
quark-antiquark bound state problem is in terms of the Coulomb potential. At  $x > \bar{x}_0(T)$, the entropic force becomes stronger than the Coulomb one, and it pulls the bound state apart, leading to the dissociation of quarkonium. In this regime, to describe the real-time dynamics of the process one would have to include the drag force on heavy quarks, which is also $\sim \sqrt{\lambda}$ \cite{Herzog:2006gh, Gubser:2006bz} -- same order in the coupling as both the Coulomb force and the entropic force (\ref{string_force}). 
\vskip0.2cm
As the temperature increases, $\bar{x}_0(T) \sim T^{-1}$ shrinks, leading to the dissociation of smaller and smaller quarkonium states -- this is consistent with the ``sequential quarkonium suppression" scenario \cite{Karsch:2005nk}. The monotonic dependence of $\bar{x}_0(T)$ on temperature is a consequence of  conformal invariance of ${\cal{N}}=4$ supersymmetric Yang-Mills theory. 
\vskip0.2cm
The entropic force (\ref{dif_force}) is suppressed relative to (\ref{string_force}) by $1/\sqrt{\lambda}$; if we include this sub-leading term in (\ref{balance}), we get a quartic equation. Since the general solution of this equation is quite cumbersome, it is convenient to fix the value of the coupling constant $\lambda$ prior to solving it. We choose again the value of the coupling $\lambda = 5.5$ determined \cite{Gubser:2006qh} from the fitting of the Coulomb potential (\ref{coul}) to the Coulomb part of the quark-antiquark potential measured in lattice QCD; in QCD, it corresponds to $\alpha_s = \lambda/(4 \pi N) \simeq 0.15$ where $N=3$ is the number of colors. With this value of the coupling, the quartic equation (\ref{balance}) has a single real and positive root, $t_0 \simeq 0.2355$. It corresponds to the following mean distance squared:
\be
\langle x_0^2 \rangle = 0.028 \ T^{-2} .
\ee
At temperature $T = 200$ MeV we get
 \be
\bar{x}_0 (T = 200\ {\rm MeV}) \simeq 0.17\ {\rm fm},
\ee 
which as expected is a little smaller than (\ref{est_strong}) and corresponds to the large contribution of the entropic force to the suppression of charmonium states at $T_c$.

\vskip0.2cm
In the confining Yang-Mills theory, the growth of the quark-antiquark entropy with the distance $L$ is even faster (\ref{Ldep}), $\sim L^4$ than $\sim L^3$ in the ${\cal{N}}=4$ supersymmetric YM case,  and thus can lead to a stronger entropic force. To compare the numerical values of the entropy in these cases, we can evaluate it at $L=0.24/T_c$ and temperature just above $T_c$ -- for the confining YM we get $S(L=0.24/T_c) \simeq 0.74$  and for the ${\cal{N}}=4$ supersymmetric YM $S(L=0.24/T_c) \simeq 0.98$. In addition, the entropy in the case of confining Yang-Mills theory possesses a sharp peak at $T_c$ shown in Fig. \ref{fig:entropy2} and this strongly enhances the entropic force at $T=T_c$. We thus conclude that around the deconfinement transition the entropic destruction of charmonium states should be very effective.

\section{Summary and discussion}
\label{sec:discussion}
\vskip0.3cm
The large amount of entropy associated with the heavy quark-antiquark pair immersed in the quark-gluon plasma 
found by the lattice QCD \cite{Kaczmarek:2002mc,Petreczky:2004pz,Kaczmarek:2005zp,Kaczmarek:2005gi} indicates a strong degree of entanglement between the pair and the rest of the system. The strong and sharp peak in the entropy of the quark-antiquark pair at the deconfinement transition is a salient and previously unexplained feature of the lattice data.
It is very likely that the quark-gluon plasma around the deconfinement transition is strongly coupled -- therefore in this paper we have used the holographic correspondence to study the entropy  associated with the heavy quark-antiquark pair in two theories: i) ${\cal{N}}=4$ supersymmetric Yang-Mills and ii) a confining Yang-Mills theory obtained by compactification on a Kaluza-Klein circle \cite{Witten:1998zw}. 
\vskip0.2cm
In both cases we have found the entropy growing with the inter-quark distance, and thus giving rise to the entropic forces that tend to destroy the bound quarkonium states. Moreover, in the case of the confining Yang-Mills theory we have found a sharp peak in the quark-antiquark entropy at the deconfinement transition, in accord with the lattice QCD. The strength of this peak in our description is within $\sim 25\%$ from the lattice value. The peak of the quark-antiquark entropy around the deconfinement transition temperature $T_c$  supports the proposal \cite{Kharzeev:2014pha} that the charmonium suppression driven by the entropic force should be the strongest in this temperature range.
\vskip0.2cm
The origin of this peak in the holographic description is intriguing -- it arises because the heavy quark pair acts as an eyewitness of the black hole formation in the confining (at low temperatures) bulk geometry. This process of black hole formation is the dual holographic representation of the deconfinement transition on the boundary. From this viewpoint, the entropy associated with the quark-antiquark pair is the right quantity to detect the temperature at which the deconfinement occurs. 
\vskip0.2cm
Our proposal of using the entropy associated with the heavy quark-antiquark pair to detect the deconfinement transition is somewhat similar to the idea \cite{Ryu:2006bv} of using the entanglement entropy as an order parameter of deconfinement. The difference is that the order parameter discussed in \cite{Ryu:2006bv} is the von Neumann entanglement entropy of a spatial region with a boundary, while we consider the Gibbs entropy associated with the quark-antiquark pair.  
In terms of the boundary theory, the entropy of the quark-antiquark pair likely emerges from the entanglement of a ``long string" \cite{Kogut:1974ag,Pisarski:1982cn,Patel:1983sc,Aharonov:1987ah,Deo:1989bv,Levin:2004mi,Kalaydzhyan:2014tfa,Hashimoto:2014xta,Hanada:2014noa} connecting the quark and antiquark with the rest of the system. 
It would be interesting to clarify this issue further as it may improve our understanding of both deconfinement and confinement.  
\vskip0.2cm
This work was supported in part by the U.S. Department of Energy under Contracts No.
DE-FG-88ER40388 and DE-AC02-98CH10886, and by the RIKEN iTHES project.

%%%%%%%%

\end{document}